\documentclass[aps,prd,preprint,nofootinbib]{revtex4}
\usepackage{epsfig}

\def\bfsigma{\mbox{\boldmath $\sigma$}}

\def\bfepsilon{\mbox{\boldmath $\epsilon$}}

\def\lQ{\Lambda_{\rm QCD}}

\def\siml{{\ \lower-1.2pt\vbox{\hbox{\rlap{$<$}\lower6pt\vbox{\hbox{$\sim$}}}}\ }}
\def\simg{{\ \lower-1.2pt\vbox{\hbox{\rlap{$>$}\lower6pt\vbox{\hbox{$\sim$}}}}\ }}

\def\vbfD{{\ \lower-8pt\vbox{\hbox{\rlap{$\!\leftrightarrow$}\lower8pt\vbox{\hbox{$\!\bf D$}}}}\ }}
\def\dsl{\,\raise.15ex\hbox{/}\mkern-13.5mu D}

\newcommand{\nn}{\nonumber}
\newcommand{\be}{\begin{equation}}
\newcommand{\ee}{\end{equation}}
\newcommand{\bea}{\begin{eqnarray}}
\newcommand{\eea}{\end{eqnarray}}
\newcommand{\beq}{\begin{equation}}
\newcommand{\eeq}{\end{equation}}
\newcommand{\bqa}{\begin{eqnarray}}
\newcommand{\eqa}{\end{eqnarray}}

\newcommand{\Appendix}[1]%
    {%
     \section{#1}%
      }

\begin{document}
\preprint{\vbox{\halign{ &# \hfil\cr &\today\cr }}}
\title{Which hadronic decay modes are good for $\eta_b$ searching:
double $J/\psi$ or something else?}
\author{Yu Jia}
\affiliation{Institute of High Energy Physics and Theoretical
Physics Center for Science Facilities, Chinese Academy of Sciences,
Beijing 100049, China}

\begin{abstract}
It has been controversial whether $\eta_b$ can be discovered in
Tevatron Run 2 through the decay  $\eta_b\to J/\psi\,J/\psi$
followed by $J/\psi\to \mu^+\mu^-$. We clear this controversy by an
explicit calculation which predicts ${\rm Br}[\eta_b\to
J/\psi\,J/\psi]$ to be of order $10^{-8}$. It is concluded that
observing $\eta_b$ through this decay mode in Tevatron Run 2 may be
rather unrealistic. The $\eta_b$ may be observed in the forthcoming
LHC experiments through the 4-lepton channel, if the background
events can be significantly reduced by imposing some kinematical
cuts. By some rough but plausible considerations, we find that the
analogous decay processes $\eta_b \to VV,\: D^*\overline{D}^*$ also
have very suppressed branching ratios, nevertheless it may be worth
looking for $\eta_b$ at LHC and Super B factory through the decay
modes $\eta_b \to K_S K^{\pm}\pi^{\mp},\:D^*\overline{D}$.
\end{abstract}


\maketitle

\newpage

\section{Introduction}

Since the discovery of $\Upsilon(1S)$ nearly three decades ago,
extensive search for its pseudo-scalar partner, $\eta_b$, has been
conducted in various experiments. Unfortunately, to date there is
still no conclusive evidence that this elusive particle has been
found~\cite{Brambilla:2004wf}.

The existence of $\eta_b$ is a solid prediction of QCD. On the
theoretical side, many work have attempted to unravel its various
properties. In particular, its mass is believed to be among the
simplest and most tractable observables. Numerous estimates for
$\Upsilon-\eta_b$ mass splitting span the range 20--140
MeV~\cite{Eichten:1994gt,
Narison:1995tw,Lengyel:2000dk,Brambilla:2001fw,
Barnes:2001rt,Liao:2001yh,Ebert:2002pp,
Recksiegel:2003fm,Kniehl:2003ap, Gray:2005ur}. Among different
theoretical approaches,  perturbative QCD is believed by many people
to yield reliable predictions,  because of decently heavy $b$ mass
and $\eta_b$ being the lowest-lying $b\overline{b}$ state. By far
the most sophisticated prediction along this direction, facilitated
by the NRQCD renormalization group technique, gives $M_{\eta_b} =
9.421 \pm 0.013$  GeV~\cite{Kniehl:2003ap}. An eventual unambiguous
observation of $\eta_b$ and precise measurement of its mass will
decisively test the weakly-coupled picture of the $\bar{b}b$ ground
state.

Much efforts are spent to search $\eta_b$ from $\gamma\gamma$
collisions in the full data samples of LEP 2, where approximately
two hundreds of $\eta_b$ are expected to be produced. ALEPH has one
candidate event $\eta_b\to K_S (\to\pi^+\pi^-) K^- \pi^+ \pi^-
\pi^+$ (possibly missing a $\pi^0$) with the reconstructed mass of
$9.30\pm 0.03$ GeV, but consistent to be a background
event~\cite{Heister:2002if}. ALEPH, L3, DELPHI have also set upper
limits on the branching fractions for the $\eta_b$ decays into 4, 6,
8 charged
particles~\cite{Heister:2002if,Levtchenko:2004ku,Abdallah:2006yg}.
Based on the 2.4 ${\rm fb}^{-1}$ data taken at the $\Upsilon(2S)$
and $\Upsilon(3S)$ resonances, CLEO has searched distinctive single
photons from hindered $M1$ transitions $\Upsilon(2S),\Upsilon(3S)\to
\eta_b\gamma$ and $\Upsilon(3S)\to h_b\pi^0,\:h_b\pi^+\pi^-$
followed by $E1$ transition $h_b\to \eta_b\gamma$, but no signals
have been seen~\cite{Artuso:2004fp}.

Hadron collider experiments provide an alternative environment  to
search for $\eta_b$. Unlike the $e^+e^-$ machines which is limited
by the low yield of $\eta_b$, hadron colliders generally possess a
much larger production cross sections for $\eta_b$, which in turn
allows for searching it through some relatively rare decay modes yet
with clean signature. However, one should bear in mind that a
noteworthy disadvantage also accompanies with hadron collision
experiment, i.e., that the corresponding background events may also
be copious, so the effectiveness of these decay modes might be
seriously discounted (Such an example may be the decay mode
$\eta_b\to \gamma\gamma$, with branching fraction $\sim 10^{-4}$,
but the combinatorial background $\gamma$ events can be enormous).

Encouraged by the large observed width of $\eta_c \to VV$($V$ stands
for light vector mesons), Braaten, Fleming and Leibovich (hereafter
BFL) have suggested that the analogous decay process $\eta_b\to
J/\psi J/\psi$, followed by both $J/\psi$ decays to muon pairs,
may be used as a very clean trigger to search for $\eta_b$ at
Tevatron Run 2~\cite{Braaten:2000cm}.
Assuming ${\rm Br}[\eta_b\to J/\psi
J/\psi]\sim 1/m_b^4$, they rescale the measured branching ratio of
$\eta_c\to \phi\phi$ by a factor of $(m_c/m_b)^4$ to
estimate~\footnote{Note  that ${\rm Br}_{\rm exp}[\eta_c\to
\phi\phi]$  has shifted from $(7.1\pm 2.8)\times 10^{-3}$ given in
2000 PDG edition~\cite{Groom:2000in}, which was quoted by BFL, to
the latest PDG value $(2.7\pm 0.9)\times
10^{-3}$~\cite{Yao:2006px}.}
\bqa
{\rm Br}[\eta_b\to J/\psi J/\psi]&=&
7\times 10^{-4\pm 1}\,.
\label{BFL:predict}
\eqa
Combining with the knowledge about the production rate of $\eta_b$
at Tevatron, BFL conclude that the prospect of observing $\eta_b$
through the $4\mu$ decay mode in Run 2 is promising.

Following this suggestion, CDF has searched for the
$\eta_b\to J/\psi J/\psi\to 4\mu$ events
in the full Run 1 data sample~\cite{Tseng:2003md}.
A small cluster of seven events are seen
in the search window, where
1.8 events are expected from background,
with the statistical significance of 2.2 $\sigma$.
A simple fit infers the cluster's
mass to be $9.445\pm 0.006\,({\rm stat})$ GeV.
If this cluster is truly due to $\eta_b$,
then the product of its production cross section
and decay branching ratio is close to the
upper end of BFL expectation.

In a recent work, Maltoni and Polosa (henceforth MP) nevertheless
argue that the BFL estimate, (\ref{BFL:predict}), may be overly
optimistic~\cite{Maltoni:2004hv}. MP suspect that the analogy
between $\eta_c\to \phi\phi$ and $\eta_b\to J/\psi J/\psi$ is only
superficial.
It is known that perturbative QCD (pQCD) framework has
difficulty to account for the large
observed widths of the $\eta_c\to VV$ decay
processes~\cite{Anselmino:1990vs,jia:1998}.
The consensus is that some nonperturbative
mechanisms should be invoked to reconcile
the discrepancy between the pQCD prediction
and the actual measurement~\cite{Feldmann:2000hs,Zhou:2005fc}.
On the other hand,  due to heavy $b$ and $c$ masses,
it is rather reasonable to expect that
$\eta_b\to J/\psi J/\psi$ can be safely tackled
within the pQCD scheme.
Therefore, the rescaling procedure used by BFL,
whose validity should reside only in the domain of pQCD,
may well be illegitimate
when taking the measured ratio of $\eta_c\to \phi\phi$ as input,
which is essentially dictated by nonperturbative dynamics.

The persuasive evidence in favor of MP's argument comes from their
explicit calculation for the inclusive decay rate of $\eta_b$ to
4-charm states,
\bqa
{\rm Br}
[\eta_b\to c\overline{c}c\overline{c}]&=&
1.8^{+2.3}_{-0.8}\times 10^{-5}\,.
\label{MP:inclusive}
\eqa
This ratio is even smaller than the lower limit of ${\rm
Br}[\eta_b\to J/\psi J/\psi]$ estimated by BFL! This is clearly at
odds with the usual thought that the exclusive decay rate should be
much smaller than the inclusive one.

One of the major concerns of this paper is to dispose of this
controversy by performing an explicit calculation. Heavy $b$ and $c$
quark masses set hard scales so that we can confidently utilize pQCD
to tackle this decay process, expecting those nonperturbative
contributions plaguing the decay $\eta_c\to VV$ play an
insignificant role here. Since each involved particle is a heavy
quarkonium, we work with the color-singlet model, in line with the
calculation done for double charmonium production at $e^+e^-$
colliders~\cite{Braaten:2002fi,Bodwin:2002fk,
Bodwin:2002kk,Liu:2002wq,Hagiwara:2003cw,Zhang:2005cha,Gong:2007db}.
It is found that at the lowest order in $\alpha_s$, retaining the
transverse momentum of $c$ inside $J/\psi$ is vital to obtain a
non-vanishing result, and consequently, the correct asymptotic
behavior of the hadronic decay branching ratio is
$\alpha_s^2\,v_c^{10} (m_c/ m_b)^8$ ($v_c$ stands for the typical
velocity of $c$ in $J/\psi$). Numerically, we predict
\bqa {\rm Br}[\eta_b\to J/\psi J/\psi] &=& (0.5-6.6)\times
10^{-8}\,, \eqa
which is much smaller than the BFL estimate, (\ref{BFL:predict}).
Simple analysis indicates that the cluster reported by
CDF~\cite{Tseng:2003md} is extremely unlikely to be affiliated with
$\eta_b$.  We further argue that, the potential of discovering
$\eta_b$ through this decay mode is gloomy  even in Run 2.

To better guide experimental search for $\eta_b$, an important issue
is to know that, besides $\eta_b\to J/\psi J/\psi$,  which
hadronic decay modes can serve as  efficient triggers to detect
$\eta_b$? Heavy $\eta_b$  mass opens a huge phase space available
for innumerable decay modes,  but at a price that the
branching ratio of each individual mode has been greatly diluted
with respect to that from $\eta_c$ decay.

At any rate, it is valuable to know which hadronic decay modes of
$\eta_b$ possess the largest branching ratios. We examine numerous
hadronic decay channels, mostly two-body ones, e.g., $\eta_b$ decays
to two light mesons, and two charmed mesons. By some crude but
plausible estimate, we find $\eta_b \to \phi\phi,\: D^*
\overline{D}^*$ also have very tiny branching ratios, of the same
order as that for $\eta_b\to J/\psi J/\psi$. In sharp contrast, the
branching ratios for $\eta_b \to D^*\overline{D}$ might be as large
as $10^{-5}$, therefore they can be used as searching modes.
Furthermore, stimulated by the experimental fact that $\eta_c$
decays to three pseudoscalars have largest branching ratios, we urge
experimentalists to look into such 3-body channels as $\eta_b
\to\:K\overline{K}\pi,\: \eta(\eta^\prime)\pi\pi,\:D
\overline{D}\pi$. The corresponding ratios are estimated to be of
order $10^{-4}$.

The remainder of the paper will proceed as follows. In
Sec.~\ref{Asymptotic}, we discuss the asymptotic behavior of ${\rm
Br}[\eta_b\to J/\psi J/\psi]$ in the limit $m_b\gg m_c\gg
\Lambda_{\rm QCD}$, and elucidate the peculiarity of this decay
process. We then present the actual calculation of the ratio in
Sec.~\ref{CSM:calculation}, employing the color-singlet model which
incorporates velocity expansion. In Sec.~\ref{ob:potent:Tev:LHC}, we
discuss the observation potential of $\eta_b$ through this mode in
Tevatron and LHC. In Sec.~\ref{other:decay:channels}, by some simple
scaling analysis, we estimate the branching ratios for $\eta_b$
decays to various final states, e.g., to two light mesons,  to three
light pseudo-scalar mesons, and to two charmed mesons. For the decay
process $\eta_b\to VV$, we also compare our estimates with those
obtained from some nonperturbative mechanism.  In Sec.~\ref{summary}
we summarize and give an outlook.

\section{Asymptotic Behavior and Unnatural decay process}
\label{Asymptotic}

\begin{figure}[tb]
\begin{center}
\includegraphics[scale=0.5]{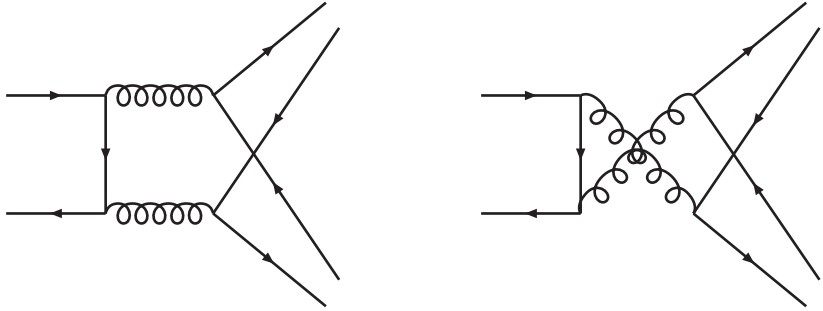}
\caption{Lowest-order QCD diagrams which contribute
to $\eta_b\to J/\psi\,J/\psi$.
\label{QCD:diagram}}
\end{center}
\end{figure}

\begin{figure}[tb]
\begin{center}
\includegraphics[scale=0.5]{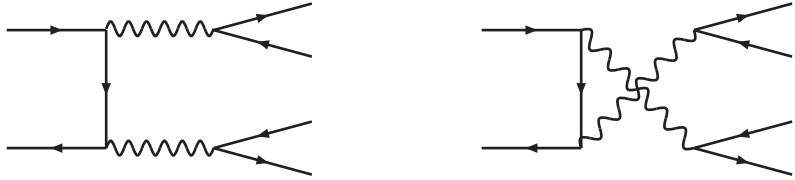}
\caption{
Lowest-order QED diagrams that contributes to
$\eta_b\to J/\psi\,J/\psi$. Only the fragmentation
type diagrams  are retained, whereas the other two,
which can be obtained by replacing the gluons in
Fig.~\ref{QCD:diagram} by photons,
have been suppressed.
\label{QED:diagram}}
\end{center}
\end{figure}

Before launching into the actual calculation, it is instructive to
first envisage the general property of this exclusive decay process.
A powerful tool in pQCD to count kinematical suppression factors for
hard exclusive processes is the so-called {\it hadron helicity
selection rule}, originally developed for the reactions involving
light hadrons~\cite{Brodsky:1981kj}. This rule has recently been
applied to analyze the double charmonium production in $e^+ e^-$
annihilation in the limit $\sqrt{s}\gg m_c$~\cite{Braaten:2002fi}.
One can work out the asymptotic behavior for ${\rm Br}[\eta_b \to
J/\psi J/\psi]$ in an analogous way. The decay  $\eta_b\to
J/\psi\,J/\psi$ can be initiated by either strong or electromagnetic
interaction, with the corresponding lowest-order diagrams shown in
Fig.~\ref{QCD:diagram} and Fig.~\ref{QED:diagram}, respectively.
Generally speaking, the QCD contribution is dominating over the QED
contribution. However, for the sake of completeness, the latter will
be also included in our analysis.

For the lowest order strong decay process depicted in
Fig.~\ref{QCD:diagram}, a simple consideration may suggest that in
the limit $m_b\to \infty$ with $m_c$ fixed,
\bqa
{\rm Br_{\rm \, str}} [\eta_b\to J/\psi(\lambda)+J/\psi(\tilde{\lambda})]
&\sim& \alpha_s^2 \,v_c^6
\left(m_c^2\over m_b^2\right)^{2+|\lambda+\tilde{\lambda}|}\,,
\label{Brstr:scaling}
\eqa
where $\lambda$ and $\tilde{\lambda}$ represent the helicities of
two $J/\psi$ viewed in the $\eta_b$ rest frame, and $v_c$ denotes
the characteristic velocity of $c$ in $J/\psi$. Obviously the
scaling behavior depends on the helicity configurations of both
$J/\psi$.  The factor $v_c^6$ is expected because there is a factor
of wave function at the origin,  $\psi_{J/\psi}(0)$, for each
$c\overline{c}$ pair to emerge with small relative momentum
to form S-wave bound state,
and $\psi_{J/\psi}(0)\sim (m_c\,v_c)^{3/2}$.

The expectation (\ref{Brstr:scaling}) is compatible with the
helicity selection rule that the decay configuration which conserves
the hadron helicity, {\it i.e.} $\lambda+\tilde{\lambda}=0$,
exhibits the slowest asymptotic decrease, ${\rm Br_{str}}\sim
1/m_b^4$. This occurs because there are two hard gluons carrying
large momentum transfers. The only helicity state bearing this least
suppressed ratio which is also compatible with the angular momentum
conservation $\lambda=\tilde{\lambda}$ is thus ($\lambda$,
$\tilde{\lambda}$)=(0, 0). The helicity conservation can be violated
either by the nonzero charm mass $m_c$ or by the transverse momentum
of $c$ inside $J/\psi$, $q_\perp$. For every unit of violation of
the selection rule, there is a further suppression factor of
$m_c^2/m_b^2$ or $q_\perp^2/m_b^2$. For the other physically allowed
configurations ($\lambda$, $\tilde{\lambda}$)=($\pm 1$, $\pm 1$),
the helicity conservation is violated by two units, so one expects
${\rm Br_{str}}\sim 1/m_b^8$. By writing (\ref{Brstr:scaling}) the
way as it is, we have temporarily assumed that the cause of
violation is entirely due to the quark mass $m_c$.

The essential assumption of BFL is the ``leading twist'' scaling
behavior ${\rm Br_{str}}\sim 1/m_b^4$, which is tacitly associated
with the $\eta_b$ decay to two longitudinally polarized
$J/\psi$~\footnote{The decay $\eta_b \to J/\psi\,\eta_c$ would
exhibit the ``leading twist" scaling behavior, ${\rm Br}\sim
\alpha_s^2 \,v_c^6 (m_c/ m_b)^4$, were it not inhibited by $C$
invariance.}. However, one may recall that $\eta_b \to J/\psi
J/\psi$, like $\eta_c\to VV$, belongs to a class of so-called {\it
unnatural} decay processes~\cite{Chernyak:1983ej}, for which the
helicity state $(0,0)$ is strictly forbidden due to the conflict
between parity and angular momentum conservation~\footnote{An
unnatural process, according to Ref.~\cite{Chernyak:1983ej}, is
defined as a heavy quarkonium two-meson decay process that does not
conserve a multiplicative quantum number called {\it naturalness},
which is defined by $\sigma=(-1)^J P$ ($J$, $P$ stand for the spin
and parity of a meson).}. In operational basis, it arises because
the decay amplitude for such a process involves the Levi-Civita
tensor and there are no enough number of independent Lorentz vectors
to contract with it for vector mesons are longitudinally polarized.
Further examples of the {\it unnatural} processes for a bottomonium
decays to two $S$-wave charmonia include $\Upsilon\:(\chi_{b2})\to
J/\psi \,\eta_c$ and $\chi_{b1}\to J/\psi \,J/\psi$, in contrast
with the {\it natural } decay processes such as $\chi_{b0,2}\to
J/\psi\, J/\psi$~\cite{Braguta:2005gw}, $\eta_c \,\eta_c$, and $h_b
\to J/\psi \,\eta_c$.

We may wish to examine this assertion closely for our case. Parity
and Lorentz invariance  constrain the decay amplitude to have the
following tensor structure:
\bqa
{\cal M}(\lambda,\tilde{\lambda}) &\propto &
\epsilon^{\mu \nu \alpha \beta}\,P_\mu \, \tilde{P}_\nu
\,\varepsilon^*_\alpha (\lambda)\,
\tilde{\varepsilon}^*_\beta(\tilde{\lambda})\,,
\label{Lorentz:tensor:structure}
\eqa
where $P$, $\tilde{P}$, $\varepsilon$ and $\tilde{\varepsilon}$ are
momenta and polarization vectors for both $J/\psi$. If both $J/\psi$
are longitudinally polarized, $\varepsilon$ and
$\tilde{\varepsilon}$ then can be expressed as linear combinations
of $P$ and $\tilde{P}$, $M({0,0})$ thus vanishes~\footnote{Exactly
with the same argument, the decay $e^+e^-\to \gamma^*\to
J/\psi(\lambda=0)+\eta_c$ turns out to be strictly forbidden.}. It
is worth emphasizing that this result is based solely on the basic
principle of  parity and angular momentum conservation, so it will
not depend on dynamical details. For example, it will be true
irrespective of whether this process is initiated by QCD or QED,
whether higher order perturbative corrections are included or not,
and whether nonperturbative QCD effects are incorporated or not.

In passing, one may notice an equivalent but more intuitive
explanation for $M({0,0})=0$ for the unnatural decay
processes~\cite{jia:1998}. It can be attributed to a peculiar
property of Clebsch-Gordon coefficients that $\langle
10|10;10\rangle=0$. We take $\eta_b\to J/\psi J/\psi$ as an example
to illustrate this. Due to parity and angular momentum conservation,
two $J/\psi$ must have the relative orbital angular momentum $L=1$
and the total spin $S=1$. However, it is impossible for the two
longitudinally polarized $J/\psi$ to couple to a $(S,\,S_z)=(1,\,0)$
state because of this vanishing Clebsch-Gordon coefficient.

Since both $J/\psi$ in the final state must be transversely
polarized, the helicity conservation is violated by two units,
Eq.~(\ref{Brstr:scaling}) then indicates $ {\rm Br_{\rm \, str}}
[\eta_b\to J/\psi_\perp\,J/\psi_\perp] \sim \alpha_s^2 \,v_c^6
\,(m_c/ m_b)^8$. Nevertheless, as the explicit calculation will
reveal, this behavior is not quite correct in counting powers of
$v_c$.  It turns out that the true asymptotic behavior is even more
suppressed,
\bqa
{\rm Br_{\rm \, str}} [\eta_b\to J/\psi(\pm 1)
+J/\psi(\pm 1)]
&\sim& \alpha_s^2 \,v_c^6
\left(m_c^2\over m_b^2\right)^{2}\,
\left(q_\perp^2\over m_b^2\right)^{2}
\sim \alpha_s^2 \,v_c^{10}
\left(m_c\over m_b\right)^8\,,
\label{Brstr:scaling:transverse}
\eqa
where $q_\perp\sim m_c v_c$ is assumed in the last term. This
implies that, at the lowest order in $\alpha_s$, the violation of
the rule should be ascribable to the nonzero transverse momentum of
$c$ in $J/\psi$, instead of the nonzero $m_c$. This is compatible
with the earlier finding that the amplitude for $\eta_c\to
V_\perp\,V_\perp$ vanishes in the collinear quark configuration,
even though the quark masses are kept
nonzero~\cite{Anselmino:1990vs}. After $q_\perp$ is included for the
light vector mesons, the decay rate then becomes nonzero, though
still too tiny to account for the measured rates~\cite{jia:1998}.

We can also infer the asymptotic behavior of the electromagnetic
contribution in Fig.~\ref{QED:diagram}. The photon fragmentation
produces transversely polarized $J/\psi$. Since the hard scale is
set by the virtuality of fragmenting photon $\sim m_c^2$, no
suppression factor $\propto 1/m_b^2$ can arise in the branching
ratio. A simple counting rule suggests that the QED fragmentation
contribution to the ratio exhibits the following  behavior
\bqa
{\rm Br_{\rm \, em}} [\eta_b \to J/\psi(\pm 1)+J/\psi(\pm 1)]
&\sim & \left({\alpha^4\over \alpha_s^2}\right)\,
v_c^6\,.
\label{Brem:scaling}
\eqa
Although the electromagnetic contribution is free of suppression  by
inverse powers of $m_b$, in general it still cannot counteract the
adversity caused by $\alpha\ll \alpha_s$.

There is also the interference term between QCD and QED
contributions, with a scaling behavior intermediate between
(\ref{Brstr:scaling:transverse}) and (\ref{Brem:scaling}).

\section{Color-Singlet Model Calculation}
\label{CSM:calculation}

In this section, we present a calculation for $\eta_b\to J/\psi
J/\psi$ in perturbative QCD scheme. As stressed before,  this scheme
is expected to generate reliable answer for this process, since the
annihilation of $b\bar{b}$ and creations of $c\bar{c}$ pairs take
place in rather short distances.  This decay process should be
contrasted with the analogous process $\eta_c\to VV$. To account for
the unnaturally large measured width of the latter, it is proposed
that some nonperturbative mechanisms, e.g., the mixing among
pseudo-scalar states $\eta-\eta^\prime-\eta_c$, either due to QCD
axial anomaly~\cite{Feldmann:2000hs}, or due to perturbative box
diagram~\cite{Zhou:2005fc}, together with quark pair creation from
vacuum, should play a prominent role. The impact of final state
interaction on the doubly-OZI-suppressed process $\eta_c\to \omega
\phi$ has also been addressed~\cite{Zhao:2006cx}. Fortunately,
neither of these nagging complications will bother us, because the
$\eta_b$ is too heavy to bear a significant mixing with $\eta_c$,
and $c$ is too heavy to easily pop out of the vacuum. The weak
inter-quarkonium Van der Vaals interaction~\cite{Peskin:1979va}
implies that the final state effects are also unimportant.

We first consider the QCD contribution in Fig.~\ref{QCD:diagram} for
$\eta_b(K)\to J/\psi (P)+J/\psi(\tilde{P})$,  where $K$, $P$ and
$\tilde{P}$ signify momenta of each quarkonium. In this work, we
will only consider the color-singlet Fock state of the quarkonium,
completely ignoring the possible color-octet effect, which is
difficult to analyze for exclusive processes in a clear-cut
way~\footnote{Note the color-singlet model can be viewed as a
truncated version of the Nonrelativistic QCD (NRQCD) factorization
approach~\cite{Bodwin:1994jh}, which still admits a factorization of
the calculable short-distance part and universal long-distance
factors in the color-singlet channel. For the exclusive processes
involving heavy quakonium, the color-single model and NRQCD  are
often used interchangeably in literature.}.

In the color-singlet model calculation, it is customary to begin
with the parton level matrix element
$b(k)\,\overline{b}(\overline{k}) \to c(p)
\,\overline{c}(\overline{p}) + c(\tilde{p})
\,\overline{c}(\overline{\tilde{p}})$, then project this matrix
element onto the corresponding color-singlet quarkonium Fock states.
It is worth stressing that because of impossibility to find a frame
such that two back-to-back fast moving $J/\psi$ become
simultaneously at rest, it is imperative to admit a manifestly
Lorentz-covariant projector.

For the $ c \overline{c}$ pair with total momentum $P$, we assign
the momentum carried by each constituent as
\bqa
p &=& {P\over 2} +q\,,
\nn \\
\overline{p}&=& {P\over 2} -q\,,
\eqa
where $q$ is the relative momentum satisfying
$P\cdot q=0$. In the rest frame of $c\overline{c}$
pair, $P^\mu$ is thus purely time-like and $q^\mu$
purely space-like.
Since the $ c \overline{c}$ pair
forms $J/\psi$, it is necessarily
in a spin-triplet color-singlet state,
and one can replace the product of the Dirac and color
spinors for $c$ and $\overline{c}$
in the final state
with the following Lorentz-covariant
projector~\cite{Bodwin:2002hg,Braaten:2002fi}:
\bqa
v(\overline{p})\,\overline{u}(p)& \longrightarrow &
\int [dq]\,{1\over 4\sqrt{2}E(E+m_c)} \,
(\not\!\overline{p} -m)
\not\!\varepsilon^*(\lambda)(\not\! P+2 E)
(\not\!p +m)
\nn \\
&\times &
 \left( {1\over \sqrt{m_c}}\phi_{J/\psi}(-q^2)\right)
\otimes
\left( {{\bf 1}_c\over \sqrt{N_c}}\right)\,.
\label{JPsi:projector}
\eqa
where $p^2=\overline{p}^2=m_c^2$,
$E=\sqrt{P^2}/2=\sqrt{m_c^2-q^2}$ is a Lorentz scalar,
and $\varepsilon^*$
is the polarization vector of $J/\psi$ satisfying
$\varepsilon(\lambda)\cdot \varepsilon^*(\lambda^\prime)
=-\delta^{\lambda\lambda^\prime}$ and $P\cdot \varepsilon^*=0$.
$N_c=3$ and ${\bf 1}_c$ stands for the unit color matrix.
The momentum space wave function $\phi_{J/\psi}(-q^2)$ is
explicitly included and the Lorentz-invariant measure
is defined as
\bqa
\int [dq] &\equiv&
\int \!\! {d^4 q\over (2\pi)^3} \,2\,E\, \delta(P\cdot q)\,.
\label{Lorentz:inv:measure}
\eqa

The introduction of the covariant projector (\ref{JPsi:projector})
unites the non-relativistic internal motion  and highly relativistic
external motion of $J/\psi$ in a coherent fashion, which is
indispensable if one wishes to systematically implement relativistic
corrections for highly energetic processes like this one. At the
final stage of the calculation, one can always choose to perform the
momentum integrals in the rest frame of each $J/\psi$ successively,
thanks to the Lorentz-invariant measure (\ref{Lorentz:inv:measure}).
We take the following integral, which arises in the zeroth order of
relativistic expansion, as an example:
\bqa
\int [dq]\, \phi_{J/\psi}(-q^2) &= &
\left. \int \!\! {d^3 q\over (2\pi)^3} \,\phi_{J/\psi}({\bf q}^2)
\right|_{rest\;\;frame}
= \psi_{J/\psi}(0) \,,
\eqa
where $\psi_{J/\psi}(0)$ is the spatial
Schr\"{o}dinger
wave function at the origin for $J/\psi$.

The above formulas  obviously also apply to the second $J/\psi$,
once the replacements  $P\to\tilde{P}$, $q\to\tilde{q}$, $E\to
\tilde{E}$, and
$\varepsilon(\lambda)\to\tilde{\varepsilon}(\tilde{\lambda})$ are
made.

For the $\eta_b$ in the initial state, it turns out that the ${\cal
O}(v_b^2)$ correction is less relevant, so we will neglect the
relative momentum and simply take $k=\overline{k}=K/2$.
Consequently, the following simplified projector will be used:
\bqa
u(k)\,\overline{v}(\overline{k})&
\longrightarrow&
{1\over 2 \sqrt{2}}\,(\not\! K+2 m_b)\,i\gamma_5
\times
\left( {1\over \sqrt{m_b}} \psi_{\eta_b}(0)\right)
\otimes
\left( {{\bf 1}_c\over \sqrt{N_c}}\right)\,.
\label{Etab:projector}
\eqa

For processes involving heavy quarkonium, one usually organizes the
amplitude in powers of the relative momenta, to accommodate the
NRQCD ansatz. Some simple algebra shows that the $\eta_b$ part of
the matrix element contains the factor
\bqa
{\rm Tr}[\gamma_5 \not\! K
\gamma^\mu (\not\!q-\not\!\tilde{q})\gamma^\nu ]
&=&  4i\,\epsilon^{\,\mu \nu \rho \sigma} K_\rho
(q-\tilde{q})_\sigma\,.
\label{etab:part:ampl}
\eqa
In the leading order in $v_c$ expansion, $q=\tilde{q}=0$, the
amplitude hence vanishes. As a matter of fact,
Anselmino {\it et al} have drew the same conclusion
for $\eta_c\to VV$ while using light-cone
scheme in the collinear quark configuration
but retaining nonzero
light quark masses~\cite{Anselmino:1990vs}.
This color-singlet-model result can be viewed
as a specific example of theirs
by invoking narrow-peak
approximation to the light-cone wave function
for vector mesons.

In order to obtain a non-vanishing amplitude, one must go to the
next-to-leading order in $v_c$. In particular,
one should expand the remaining  part of amplitude
to linear order in $q$ and $\tilde{q}$, to pair up
with the ones in
Eq.~(\ref{etab:part:ampl})~\footnote{Retaining the relative momentum
in $\eta_b$  but not $q$ and $\tilde{q}$ still leads to vanishing
amplitude.}.

First we expand the product of three propagators
in Fig.~\ref{QCD:diagram}:
\bqa
& & {1\over (q-\tilde{q})^2-m_b^2}\,{1\over (K/2-q+\tilde{q})^2}\,
{1\over (K/2+q-\tilde{q})^2}
\nn \\
 & \approx & -{1\over m_b^6}\,
\left[1+{\cal O}\left({q^2\over m_b^2},
{\tilde{q}^2 \over m_b^2}, {q\cdot \tilde{q} \over m_b^2} \right)
\right]\,.
\eqa
Because the sub-leading terms are at least quadratic in $q$ or
$\tilde{q}$, they can be dropped.

The missing $q$ or $\tilde{q}$ thus must arise from the $J/\psi$
part of the matrix element, or more precisely, from the projectors
for  $J/\psi$, Eq.~(\ref{JPsi:projector}). After some
straightforward Dirac trace algebra, one can express the full QCD
amplitude in the Lorentz-invariant form:
\bqa
{\cal M}_{\rm str} &=&
-4\sqrt{2} \,C_{\rm str} \,g_s^4 \,{\psi_{\eta_b}(0) \over m_b^{13/2} m_c }
\,A\,,
\eqa
where the color factor
$C_{\rm str}=N_c^{-3/2} {\rm Tr}(T^aT^b)\,{\rm Tr}(T^aT^b)
= 4^{-1}N_c^{-3/2}(N_c^2-1)$, and the double integral $A$ reads
\bqa
A &=& \int\!\!\!\int [dq]\, [d \tilde{q}]\,
\phi_{J/\psi}(-q^2)\, \phi_{J/\psi}(-\tilde{q}^2)\,
\nn \\
&\times &
\left[\epsilon_{\mu \nu \alpha \beta}
P^\mu \tilde{P}^\nu \,q^\alpha \,(\tilde{\varepsilon}^{*\beta}
\,q \cdot \varepsilon^*- \varepsilon^{*\beta} \,
q \cdot \tilde{\varepsilon}^*) + (q\leftrightarrow \tilde{q}) \right]
\,,
\label{Definition:double:integral}
\eqa
where only the terms with the intended accuracy of ${\cal O}(v_c^2)$
are kept. We have substituted $E,\,\tilde{E}$ everywhere by $m_c$,
which is legitimate in the current level of accuracy~\footnote{For
the same reason, we have not bothered to include explicitly the
relativistic effects due to 2-body phase space and normalization of
$c\bar{c}$ states as considered in
Ref.~\cite{Bodwin:2002hg,Braaten:2002fi}, which are powers of
$E/m_c$ or $\tilde{E}/m_c$.}.

One immediately observes that, to survive when contracted with the
antisymmetric tensor, $q$ and $\tilde{q}$, as well as
$\varepsilon^*$ and $\tilde{\varepsilon}^*$ in
(\ref{Definition:double:integral}), must be {\it transverse}. We
thus confirm the earlier assertion that transverse momentum is the
agent to violate the helicity conservation.

Because the transverse components of a 4-vector are invariant under
the boost along the moving direction of $J/\psi$,  we can perform
the integrals in the rest frame of $J/\psi$ without concerning about
the boost effect. Using the spherical symmetry of $S$-wave wave
function, we have
\bqa
\int\! [dq] \,
q_\perp^i q_\perp^j \, \phi_{J/\psi}(-q^2)
&= & {\delta^{ij}\over 3} \int \!\! \left. {d^3 q\over (2\pi)^3} \,
{\bf q}^2 \, \phi_{J/\psi}({\bf q}^2) \right|_{rest\;\;frame}
\nn \\
&\equiv & {\delta^{ij}\over 3}\,m_c^2 \,\langle v^2 \rangle_{J/\psi}
\,\psi_{J/\psi}(0)\,.
\label{transverse:integral}
\eqa
Here $\langle v^2 \rangle_{J/\psi}$ is a quantity governing the size
of relativistic corrections. Loosely speaking, it is related to the
second derivative of wave function at the origin for $J/\psi$, and
characterizes the average $v^2$ of $c$ inside $J/\psi$. Inspecting
Eq.~(\ref{transverse:integral}), however, one immediately realizes
at large $\bf q$,  the Coulomb wave function should dominate and
this quantity turns out to be linearly ultraviolet divergent, hence
its meaning becomes obscure.  In fact, this factor admits a more
rigorous definition as a ratio of NRQCD matrix
elements~\cite{Bodwin:1994jh,Braaten:2002fi}:
\bqa \langle v^2 \rangle_{J/\psi} &=& {\bfepsilon \cdot \langle
 J/\psi(\bfepsilon)|\psi^\dagger \bfsigma (-{\bf D}^2) \chi |0 \rangle \over m_c^2 \,
\bfepsilon \cdot \langle J/\psi(\bfepsilon)|\psi^\dagger \bfsigma
\chi |0 \rangle }, \eqa
where $\psi$ and $\chi$  represent Pauli spinor fields in NRQCD,
$\mathbf D$ is spatial covariant derivative. The matrix elements
appearing in the above ratio should be understood to be the
renormalized ones.  Lattice QCD extraction of this quantity has been
available long ago, but the precision is quite
poor~\cite{Bodwin:1996tg}. We will specify our choice for the
numerical value of $\langle v^2 \rangle_{J/\psi}$ in Section
\ref{ob:potent:Tev:LHC}.

Substituting (\ref{transverse:integral})
into (\ref{Definition:double:integral}), we obtain
\bqa A &=& -{4 \over 3}\, m_c^2 \, \langle v^2 \rangle_{J/\psi}\,
\psi^2_{J/\psi}(0)\, \epsilon_{\mu \nu \alpha \beta} P^\mu
\tilde{P}^\nu \,\varepsilon^{*\alpha} \,
\tilde{\varepsilon}^{*\beta} \,, \eqa
which has the desired tensor structure as in
(\ref{Lorentz:tensor:structure}).
The strong decay amplitude then reads
\bqa
{\cal M}_{\rm str} &=& {512 \sqrt{6}\, \pi^2 \,\alpha_s^2
\,m_c \over 27 \,m_b^{13/2} } \psi_{\eta_b}(0)
\, \psi_{J/\psi}^2(0) \,\langle v^2\rangle_{J/\psi}\,
\epsilon^{\mu \nu \alpha \beta}\,P_\mu \, \tilde{P}_\nu
\,\varepsilon^*_\alpha \,
\tilde{\varepsilon}^*_\beta\,.
\label{amplitude:strong}
\eqa

Next we turn to the electromagnetic contribution to $\eta_b\to
J/\psi J/\psi$. Two QED diagrams which have the same topology as
Fig.~\ref{QCD:diagram}, but with gluons replaced by photons, lead to
the amplitude of the same form as (\ref{amplitude:strong}) except
$\alpha_s^2$ is replaced by $e_b^2\,e_c^2\alpha^2$. Obviously, their
contributions are much more suppressed than those from the
fragmentation diagrams in Fig.~\ref{QED:diagram}, hence will not be
considered. The QED fragmentation contribution to the amplitude can
be easily worked out,
\bqa
{\cal M}_{\rm em} &=& {24\sqrt{6} \,
\pi^2 \,e_b^2\,e_c^2 \alpha^2\over  m_b^{5/2} \, m_c^3 }
\psi_{\eta_b}(0) \,\psi_{J/\psi}^2(0)
\,\left(1-{2 m_c^2\over m_b^2}\right)^{-1}\,
\epsilon^{\mu \nu \alpha \beta}\,P_\mu \, \tilde{P}_\nu
\,\varepsilon^*_\alpha \,
\tilde{\varepsilon}^*_\beta\,.
\label{amplitude:em}
\eqa

Adding (\ref{amplitude:strong}) and (\ref{amplitude:em}) together,
squaring, summing over transverse polarizations of both $J/\psi$ and
integrating over half of the phase space, we then obtain the partial
width $\Gamma[\eta_b\to J/\psi J/\psi]$. Nevertheless, it is more
convenient to have a direct expression for the branching ratio,
where $\psi_{\eta_b}(0)$ drops out,
\bqa
{\rm Br}[\eta_b\to J/\psi\,J/\psi]&=&
{2^{13} \,\pi^2 \,\alpha_s^2
\over 3^4 } \,{m_c^2 \over m_b^8}
\,\psi^4_{J/\psi}(0)\,
\left(1-{4 m_c^2\over m_b^2}\right)^{3/2}
\nn \\
&\times & \left[\langle v^2 \rangle_{J/\psi}\,
+\left({ 9 \,e_b\, e_c\,\alpha \over 8\,\alpha_s}
\, {m_b^2 \over m_c^2}\right)^2
\left(1-{2 m_c^2\over m_b^2}\right)^{-1}  \right]^2\,.
\label{BR:Etab:psipsi}
\eqa
In deriving this, we have approximated the
total width of $\eta_b$ by its gluonic width:
\bqa \Gamma_{\rm tot}[\eta_b] &\approx & \Gamma[\eta_b\to g g] = {8
\pi \alpha_s^2 \over 3\, m_b^2} \,\psi_{\eta_b}^2(0)\,, \eqa
where the LO expression in $\alpha_s$ and $v_b$
is used  for simplicity.

Equation (\ref{BR:Etab:psipsi}) constitutes the main formula of this
paper. One  can readily confirm the asymptotic behavior of QCD and
QED contributions  first given in Sec.~\ref{Asymptotic},
Eq.~(\ref{Brstr:scaling:transverse}) and (\ref{Brem:scaling}).

\section{Observation potential of $\eta_b\to J/\psi\,J/\psi$
at Tevatron and LHC}
\label{ob:potent:Tev:LHC}

We now explore the phenomenological implication of
(\ref{BR:Etab:psipsi}).
The input parameters are $m_b$, $m_c$, $\alpha$, $\alpha_s$,
$\psi_{J/\psi}(0)$ and $\langle v^2\rangle_{J/\psi}$,
all of which can be inferred from other
independent sources.
The wave function at the origin for $J/\psi$ can be
extracted from its dielectron width:
\bqa
\Gamma[J/\psi\to e^+e^-] &=&
{4 \pi e_c^2 \alpha^2 \over m_c^2} \psi_{J/\psi}^2(0)\,.
\label{onium:ee:width}
\eqa
(The LO formula in $\alpha_s$ and $v_c^2$ is used for simplicity.)
Using the measured dielectron width 5.55 keV~\cite{Yao:2006px}, we
obtain $\psi_{J/\psi}(0)=0.205$ ${\rm GeV}^{3/2}$ for $m_c=1.5$
GeV.

Among various input parameters, the least precisely known is
$\langle v^2\rangle_{J/\psi}$.  Since this is a subtracted quantity,
it can be either positive or negative.  There is a useful relation,
first derived by Gremm and Kapustin using equation of motion of
NRQCD~\cite{Gremm:1997dq}, relating this quantity with the pole mass
of charm quark and the $J/\psi$ mass~\footnote{For a reformulation
of Gremm-Kapustin relation from a even lower energy effective theory
of NRQCD, dubbed {\it potential} NRQCD,  we refer the interested
readers to Ref.~\cite{Brambilla:2002nu}.}.  To our purpose this
relation reads~\cite{Braaten:2002fi}
\bqa \langle v^2\rangle_{J/\psi} & \approx & {M_{J/\psi}^2-4\,
m^2_{c\, {\rm pole}}\over 4\, m^2_{c\,{\rm pole}}}\,. \eqa
In the analysis of the process $e^+e^-\to J/\psi \eta_c$ at $B$
factory~\cite{Braaten:2002fi}, the charm quark pole mass is taken as
the commonly quoted value, 1.4 GeV, consequently $\langle
v^2\rangle_{J/\psi}=0.22$ and $\langle v^2\rangle_{\eta_c}=0.13$.
However, since  $m_{c\, {\rm pole}}$ contains renormalon ambiguity,
it cannot be determined better within an accuracy of order
$\Lambda_{\rm QCD}$. Therefore one should not be surprised that
somewhat larger value of $m_{c\, {\rm pole}}$ may be occasionally
reported in literature. For example, a recent QCD moment sum rule
analysis claims $m_{c\,{\rm pole}}=1.75 \pm 0.15$
GeV~\cite{Eidemuller:2002wk}.   For this specific value of charm
quark pole mass, one would obtain instead $\langle v^2
\rangle_{J/\psi}=-0.22\pm 0.15$.

The situation seems to be rather obscure since even the sign of
$\langle v^2 \rangle_{J/\psi}$ cannot be unambiguously determined.
To proceed, we take a practical attitude and adopt the value
$\langle v^2\rangle_{J/\psi}=0.25 \pm 0.09$ from a recent Cornell
potential model based analysis~\cite{Bodwin:2006dn}.  This {\it
positive} value seems to be most welcomed to alleviate the alarming
discrepancy between the predicted and the measured cross section for
$e^+e^- \to J/\psi+\eta_c$~\cite{Braaten:2002fi},  also seems
compatible with a recent QCD sum rule determination of $\langle
v^2\rangle_{\eta_c}$~\cite{Braguta:2006wr}.  However, we leave the
possibility open that  this quantity may turn out to be {\it
negative} after future scrutiny. Taking $m_b =M_{\eta_b}/2 \approx
4.7$ GeV, $m_c=1.5$ GeV, $\alpha=1/137$ and $\alpha_s(m_b)=0.22$, we
then find~\footnote{Note the actual prediction, ${\rm Br}\sim
10^{-8}$, is much larger than the expectation based on the
asymptotic scaling behavior ${\rm Br}\sim \alpha_s^2 \,v_c^{10}
(m_c/ m_b)^8 \sim 10^{-11}$. This can be attributed to the large
prefactors in the right hand side of (\ref{BR:Etab:psipsi}).
\label{prefactor}}
\bqa
{\rm Br}[\eta_b\to J/\psi\, J/\psi] &= & 2.4^{+4.2}_{-1.9}\times 10^{-8}\,.
\label{etab:to:JpsiJpsi:prediction}
\eqa
The uncertainty is estimated by varying $m_b$ and $m_c$ in the
$\pm 100$ MeV range, varying $\alpha_s(\mu)$ in the $-0.04$ range
(which  corresponds to slide the scale from $\mu=m_b$ to $2m_b$),
as well as taking into account the errors in measured
$\Gamma_{e^+e^-}$ (of $\pm 0.14$ keV) and in $\langle
v^2\rangle_{J/\psi}$. The constructive interference between
electromagnetic and strong amplitudes has modest effect, i.e.,
neglecting the QED contribution decreases the branching ratio by
few to ten per cents.

Our prediction for the branching ratio is at least three
order-of-magnitudes smaller than the BFL estimate,
(\ref{BFL:predict}).  Despite the large uncertainties inherent in
various input parameters, we believe our prediction captures the
correct order of magnitude, $10^{-8}$. It is also worth noting
that our prediction for this exclusive decay ratio is about one
thousandth of the inclusive 4-charm ratio, (\ref{MP:inclusive}),
which seems fairly reasonable.

Experimentally $J/\psi$ can be tagged cleanly by its decay to muon
pair. Multiplying (\ref{etab:to:JpsiJpsi:prediction}) by the
branching ratios of 6\% for each of the decay $J/\psi\to \mu^+
\mu^-$, we obtain $ {\rm Br}[\eta_b\to J/\psi J/\psi \to 4 \mu]
\approx (0.2-2.4)\times 10^{-10}$. The total cross section for
$\eta_b$ production at Tevatron energy is about 2.5 $\mu{\rm
b}$~\cite{Maltoni:2004hv}. Therefore, the production cross section
for this 4$\mu$ decay mode is about $0.05-0.6$ fb. For the full
Tevatron Run 1 data of 100 ${\rm pb}^{-1}$, we then obtain between
0.005 and 0.06 produced events. We now can safely assert that the
seven $4\mu$  events reported  by Ref.~\cite{Tseng:2003md} must
come from sources other than $\eta_b$ decay.

Tevatron Run 2 plans to achieve an integrated luminosity of 8.5
${\rm fb}^{-1}$ by 2009. Assuming equal $\sigma(p\overline{p}\to
\eta_b +X)$ at $\sqrt{s}=1.96$ and $1.8$ TeV, we then estimate
there are about 0.4-5 produced events.  Since the kinematical
cuts, as well as taking into account the acceptance and efficiency
for detecting muon, will further cut down this number, we conclude
it is not realistic to search $\eta_b$ through this decay mode in
Tevatron Run 2.

To fathom the observability of $\eta_b$ through this mode at LHC,
we need first know the inclusive $\eta_b$ production rate. There
are rough estimates for the $\chi_{b0,2}$ cross sections at LHC,
which are about 6 times larger than the corresponding cross
sections at Tevatron~\cite{Braguta:2005gw}. Assuming the same
scaling also holds for $\eta_b$, we then expect the cross section
for $\eta_b$ at LHC to be about 15 ${\rm \mu b}$,  subsequently
the production cross section for the 4 $\mu$ events to be about
0.3-3.6 fb.  For a 300 ${\rm fb}^{-1}$ data, which is expected to
be accumulated in one year running at LHC design luminosity, the
number of produced events may reach between 100 to 1000. The
product of acceptance and efficiency for detecting $J/\psi$ decay
to $\mu^+\mu^-$ is estimated to be $\epsilon\approx
0.1$\cite{Braaten:2000cm}, which is perhaps a conservative
estimate for LHC.  Multiplying the number of the produced events
by $\epsilon^2$, we expect between 1 and 10 observed events per
year.  If we loose the constraint that $J/\psi$ must be tagged by
$\mu^+\mu^-$ pair and also allow its reconstruction through
$e^+e^-$ mode, we can have 4-40 observed 4-lepton events per year.

The above analysis seems to indicate that, the chance of observing
$\eta_b$ at LHC through the 4-lepton mode subsists, but critically
hinges on whether the signal events can be singled out from the
abundant background events. The most important background events
may come from the direct double $J/\psi$ production from $gg$
fusion~\cite{Barger:1995vx, Qiao:2002rh}. From previous analysis,
we know that all of the seven $4\,\mu$ candidate events selected
by CDF based on Tevatron Run 1 data~\cite{Tseng:2003md}, should be
regarded as this kind of background events,  which seem to
outnumber the expected signal events by several orders of
magnitude.  The same situation may also apply to LHC. It might be
possible for experimentalists to judiciously choose kinematical
cuts to significantly suppress the background events while
retaining as many signal events as possible.

\section{Other decay channels of $\eta_b$}
\label{other:decay:channels}

The decay channel we have considered so far,  $\eta_b\to J/\psi
J/\psi$, which has very clean signature, is unfortunately very
much suppressed because of its maximal violation of the helicity
selection rule\footnote{This process is the only possible one for
$\eta_b$ decays to two ground charmonium states allowed by $C$ and
$P$ invariance.}. There are other two-body decay channels, e.g.,
$\eta_b$ decays to two light mesons and to two charmed mesons,
some of which do conserve the hadron helicity, thus may have much
larger branching ratios.  In this section, we attempt to estimate
the order of magnitude of the branching ratios for these
processes.  In addition, we also present some crude estimation for
the $\eta_b$ decays to three pseudo-scalar states.

\subsection{$\eta_b$ ($\eta_c$) decays
to two and three light mesons} \label{etab:2:3:mesons}

When contending with light mesons in hard exclusive processes, the
most appropriate description of them is in terms of light-cone
expansion approach.  On the other hand, the constituent quark model,
which treats the light mesons as non-relativistic bound states, is
also frequently invoked as an alternative method for a quick
order-of-magnitude estimate.  In this sense, the preceding formulas
derived for $\eta_b\to J/\psi J/\psi$ can be applied to describe the
unnatural decay processes $\eta_b \to VV$,  once we understand we
are working with the constituent quark model.

We take $\eta_b\to \phi\phi$ as an representative. By regarding
$\phi$ as a strangeonium,  we can directly use
(\ref{BR:Etab:psipsi}),  only with some trivial changes of input
parameters.
 We take the constituent quark mass $m_s \approx
M_\phi/2 = 0.5$ GeV. The wave function at the origin of $\phi$,
$\psi_{\phi}(0)$, can be extracted analogously from its measured
dielectron width of $1.27\pm 0.04$ keV~\cite{Yao:2006px} through
(\ref{onium:ee:width}). We take $\langle v^2\rangle_\phi \approx
1$ to reflect the fact that $s$ is inherently not a heavy quark
and $\phi$ is not truly a heavy quarkonium.  Taking $m_b=4.7\pm
0.1$ GeV,  varying the strong coupling constant between
$\alpha_s(m_b)=0.22$ and $\alpha_s(2m_b)=0.18$, and
including the
experimental uncertainty in $\Gamma_{e^+e^-}$, we obtain
\bqa {\rm Br}[\eta_b\to \phi\phi] &= & (0.9-1.4)\times 10^{-9}\,.
\label{etabphiphi:pQCD:CQM} \eqa
The interference effect is more pronounced in this case due to
larger ratio of $m_b$ to $m_s$. Neglecting QED contribution will
decreases the branching fraction by about 20\%. If this estimate is
reliable, such a rare decay mode perhaps will never be observed
experimentally.

It is interesting also to consider the similar decay process $\eta_c
\to \phi\phi$, which has been measured long ago.  Parallel to the
preceding procedure,  varying the strong coupling constant from
$\alpha_s(m_c)=0.36$ to $\alpha_s(2m_c)=0.26$,  taking $m_c=1.5\pm
0.1$ GeV, we obtain
\bqa {\rm Br}[\eta_c\to \phi\phi] &=& (0.3-1.5)\times 10^{-5}\,.
\label{etac:phi:phi:pQCD} \eqa
The interference effect is rather modest, {\it i.e.}, neglecting QED
contribution only decreases the branching ratio by less than one per
cent. This prediction is consistent with an early constituent quark
model based analysis using Bethe-Salpeter bound state
formalism~\cite{jia:1998}~\footnote{The predictions of
Ref.~\cite{jia:1998}, which employ three different forms of BS wave
functions for $V$, are duplicated in Table 1 of
Ref.~\cite{Ablikim:2005yi}.}. Note this estimate is indeed much
smaller than the measured value ${\rm Br}_{\rm exp}[\eta_c\to
\phi\phi]=(2.7\pm 0.9)\times 10^{-3}$, which reflects a generic
symptom in pQCD calculation,  also present in the $\eta_c$ decays to
$\rho\rho$ and $K^*\overline{K}^*$.

As stressed several times before, some nonperturbative mechanisms
must be called for to rescue this discrepancy.  Among different
proposals,  a particularly attractive and predictive scheme has been
put forward by Feldmann and Kroll (henceforth
FK)~\cite{Feldmann:2000hs}, by generalizing their influential work
together with Stech on $\eta-\eta'$ mixing in quark flavor
basis~\cite{Feldmann:1998vh}, to include $\eta_c$. FK models the
amplitude of $\eta_c\to VV$ as the product of a factor governing the
small admixture amplitude between $\eta_c$ and $\eta^{(\prime)}$,
which presumably can be inferred from QCD $U_A(1)$ anomaly,  times a
soft vertex function parameterizing the amplitude for virtual
$\eta,\,\eta'$ transiting into $VV$.   To be specific, the
$\eta_c\to \phi\phi$ in their ansatz  can be described as
\bqa \Gamma_{\rm FK} [\eta_c\to \phi\phi] &= & {1\over 32\pi
M_{\eta_c}} \left(1-{4M_\phi^2 \over M_{\eta_c}^2}\right)^{1/2}
\left|c_{\rm \phi\phi}^{\rm mix} \, g^{\rm mix}(\eta_c)\right|^2\,,
\eqa
where the subscript FK is used to distinguish from our earlier
prediction based on the pQCD  plus constitute quark model ansatz,
Eq.~(\ref{etac:phi:phi:pQCD}).   The parameter $c^{\rm mix}$ depends
on the specific flavor content of $VV$ states but not on the initial
charmonium state,  whereas the charm mass dependence is encoded in
\bqa g^{\rm mix}(\eta_c) &=& {1\over
f_{\eta_c}}\,F_{\eta^{(\prime)}VV} (s=M^2_{\eta_c})= {1\over
f_{\eta_c}}\,F_{\eta^{(\prime)} VV}(s=0)\left( {\Lambda^2\over
M^2_{\eta_c}-\Lambda^2}\right)^n, \eqa
where $f_{\eta_c}$ is the decay constant of $\eta_c$, the second
factor indicates the on-shell coupling $\eta^{(\prime)} VV$, and the
last factor parameterizes the $s-$dependence of this vertex
function. $\Lambda$ is the cutoff of typical size around 1 GeV, and
the variable $n=1$ corresponds to the familiar monopole ansatz for
form factor, and $n=2$ characterizes the dipole ansatz. One
noteworthy fact is that, this simple model seems able to account for
both the absolute and the relative strengths of partial widths for
each decay channel ($VV=\rho\rho, K^*K^*$ and $\phi\phi$), to a
reasonably satisfactory degree.

One curious question is to ask,  whether the FK mixing mechanism can
be extrapolated to $\eta_b\to VV$, and if so, whether the
corresponding prediction to the  branching ratios differs
drastically from our pQCD-based estimate. Let us now examine this
question.   After some straightforward derivation, it is easy to
find
\bqa {{\rm Br}_{\rm FK} [\eta_b\to \phi\phi] \over {\rm Br}_{\rm FK}
[\eta_c\to \phi\phi]} &\approx &  \left(1-{4M_\phi^2 \over
M_{\eta_c}^2}\right)^{-1/2} \left( {f_{\eta_c} M^n_{\eta_c}\over
f_{\eta_b} M^n_{\eta_b} }\right)^4 = 0.16\times 0.01^n,
\label{ratio:Br:FK}
 \eqa
where we have resorted to heavy quark spin symmetry for decay
constants $f_{\eta_c}\approx f_{J/\psi}$ and $f_{\eta_b}\approx
f_{\Upsilon}$, and consequently taken $f_{\eta_c}=417$ MeV and
$f_{\eta_b}=715$ MeV through the relation
$f_{J/\psi}=\sqrt{12/M_{J/\psi}}\,\psi_{J/\psi}(0)$.  For
simplicity, we have utilized the hierarchy $m_b,m_c\gg \Lambda$ in
deriving (\ref{ratio:Br:FK}). If ${\rm Br}_{\rm FK} [\eta_c\to
\phi\phi]$ is identified with the  measured value $2.7\times
10^{-3}$, we then obtain  ${\rm Br}_{\rm FK} [\eta_b\to \phi\phi]$
to be about $4\times10^{-6}$ for $n=1$ and $4\times 10^{-8}$ for
$n=2$. It is easy to see that the ${\rm Br}_{\rm FK} [\eta_b\to VV]$
scales as $m_b^{-4(1+n)} v_b^{-6}$,  since $f_{\eta_b}\sim m_b
v_b^{3/2}$. This indicates this nonperturbative mechanism, also
resemble the pQCD analysis in that the branching ratio is highly
suppressed with respect to ``leading twist" scaling $\propto
1/m_b^4$.   Despite large uncertainty, the predicted branching
ratios of $\eta_b\to VV$ in this mixing ansatz are (much) larger
than our earlier predictions based on ``NRQCD" plus constitute quark
model, (\ref{etabphiphi:pQCD:CQM}). It is reasonable to question the
validity of the above nonperturbative mechanism at $\eta_b$ energy,
in our opinion, perhaps the pQCD result in
(\ref{etabphiphi:pQCD:CQM}) is more trustworthy. At any rate, it
will be of interest for future experiments to distinguish these
different mechanisms, though the task of recording this rare decay
mode in hadron machine could be extremely challenging.

One related class of decay channels may deserve some attention. The
helicity-conserving decay, such as  $\eta_c \to K^*_\parallel
\overline{K}$ (``$\parallel$" implies longitudinally-polarized), has
still not been observed yet. The experimental bound for this channel
is~\cite{Yao:2006px}~\footnote{Although this decay channel is
favored by helicity selection rule, it violates the U-spin
conservation~\cite{Brambilla:2004wf,Feldmann:2000hs}.}
\bqa
{\rm Br_{exp}}[\eta_c\to K^* \overline{K}+ {\rm c.c.}]
&<& 1.28\,\%\,.
\label{etac:Kstar:K}
\eqa

Since the coupling $\eta^{(\prime)}VP$ is negligible, it is
reasonable to assume that the aforementioned nonperturbative
mechanism responsible for $\eta_c\to VV$,  may not play a
significant role for $\eta_c\to VP$.  One may then make use of the
scaling behaviors derived earlier to interconnect each of them.
Firstly we expect that flavor $SU(3)$ is respected in pQCD
calculation of $\eta_c\to VV$,  accordingly our prediction is ${\rm
Br}[\eta_c\to K^* \overline{K}^*]\sim {\rm Br}[\eta_c\to
\phi\phi]\sim 10^{-5}$. As we have learned from
Sec.~\ref{Asymptotic},  at tree level, the maximal helicity
violation in $\eta_b\to J/\psi J/\psi$ is entirely due to the charm
quark transverse momentum, as is evident in
(\ref{Brstr:scaling:transverse}). We assume the similar pattern
occurring here, {\it i.e.}, the branching ratios of $\eta_c \to VV$
are also suppressed by $(q_\perp/m_c)^4$ relative to those of the
corresponding helicity-conserving channels. We then have
\bqa {\rm Br}[\eta_c\to K^* \overline{K}] &\sim & {\rm Br}[\eta_c\to
\phi \phi] \left( {m_c\over \Lambda_{\rm QCD}} \right)^4
\epsilon^2_{\rm SU(3)} \sim 10^{-3}\times \epsilon^2_{\rm SU(3)}\,,
\eqa
where we have taken $q_\perp\sim \Lambda_{\rm QCD}= 500$ MeV. We
have intentionally included a parameter $\epsilon_{\rm SU(3)}$, to
embody the extent of SU(3) flavor violation at the amplitude level.
Obviously, this parameter should be proportional to current quark
mass difference $m_s-m_{d,u}$,  divided by some typical hadronic
scale (presumably independent of $m_c$).  In contrast to the isospin
violation effect, the U-spin violating process should not receive
too severe suppression.  If assuming somewhat optimistic value
$\epsilon_{\rm SU(3)}\approx 0.3$, we then obtain ${\rm
Br}[\eta_c\to K^* \overline{K}]\sim 10^{-4}$, which is safely below
the experimental bound, (\ref{etac:Kstar:K}).  Needless to say, it
will be very beneficial for the ongoing high-luminosity charmonium
facility such as BES III to impose more tight constraint on this
branching ratio.

Carrying over the same line of argument to $\eta_b$ decay to $VP$,
we obtain
\bqa {\rm Br}[\eta_b\to K^* \overline{K}] &\sim & {\rm Br}[\eta_b\to
\phi \phi] \left( {m_b\over \Lambda_{\rm QCD}} \right)^4
\epsilon^2_{\rm SU(3)} \sim 10^{-5} \times \epsilon^2_{\rm SU(3)}\,.
\eqa
If we again take $\epsilon^2_{\rm SU(3)}\approx 0.1$,  the branching
ratio is estimated to be around $10^{-6}$,  about one hundred times
larger than ${\rm Br}[\eta_b\to J/\psi\,J/\psi]$.  Of course, one
should not take this crude estimate too seriously,  and a concrete
pQCD calculation based on light-cone expansion approach, which
incorporates $m_s$ and $m_d$ difference, might be illuminating. If
this estimate is trustworthy,  one then expects there are already
about $O(10^2)$ produced events in the Tevatron Run 1.  Since $K^*$
almost exclusively decays to $K\,\pi$, one needs to select those
resonant $K\overline{K}\pi$ events. However, there are practical
problems about usefulness of this kind of decay mode in hadron
collision experiments. Because of copiously produced kaons and pions
in a typical hadron collision, huge combinatorial backgrounds might
make it very difficult to identify the true signal. On the other
hand, the prospective Super $B$ factory, which runs at several
$\Upsilon(nS)$ resonances with an unprecedented
luminosity~\cite{Hewett:2004tv}, may produce an enormous amount of
$\eta_b$ through $M1$ transition from $\Upsilon(nS)$ states. With
much suppressed backgrounds, Super $B$ factory may offer a viable
environment to detect this decay mode.

Finally we exploit one experimental fact that some
3-body channels, {\it i.e.},
$\eta_c$  decays to three pseudo-scalar states,
have exceedingly large branching ratios.
To be concrete, three $\eta_c$
decay  modes with largest branching ratios are~\cite{Yao:2006px}
\bqa {\rm Br}_{\rm exp}[\eta_c\to K \overline{K}\pi] &= & (7.0\pm
1.2)\% \,.
\\
{\rm Br}_{\rm exp}[\eta_c\to \eta \pi\pi] &=& (4.9\pm 1.7)\% \,.
\\
{\rm Br}_{\rm exp}[\eta_c\to \eta^\prime \pi\pi] &= & (4.1\pm 1.8)\%
\,. \eqa
Although we do not know how to calculate these processes in
practice,  some general pattern may still be identified. Subtracting
off the phase space effects, one finds these three amplitudes
roughly respect the  SU(3) flavor symmetry. This may be taken as the
sign that these processes could in principle be described  by the
pQCD scheme. These processes proceed as two steps. Firstly $\eta_c$
annihilates to two highly virtual gluons, which then transit into
four energetic light quarks. This is a short-distance process.
Subsequently these light quarks materialize into three
pseudo-scalars via soft nonpertubative dynamics, which is a
long-distance process. Factorization is expected to hold between
these two stages due to the hard scale set by large $c$ mass.
Because of their large decay ratios, these processes are naturally
expected to possess the ``leading twist" scaling behaviors, ${\rm
Br}\sim 1/m_c^4$. This scaling assumption can then be used to infer
the ratios for $\eta_b$ decays to the same pseudo-scalar states. For
example, we may expect
\bqa
{\rm Br}[\eta_b\to K \overline{K}\pi] & \sim & {\rm Br}_{\rm
exp}[\eta_c\to K \overline{K}\pi] \left( {m_c \over m_b} \right)^4
\sim 10^{-4} \,.
\eqa
This is by far the largest branching ratio we have found
in all exclusive hadronic decay channels of $\eta_b$.
If this is the case, these channels will be worth pursuing.
Practically, the $\eta \pi\pi$ mode may not be
very useful in hadron hadron collider experiments,
since the ordinary  way of reconstructing $\eta$,
which goes through the $2\gamma$ decay,  may suffer
enormous contamination from combinatorial background.
For similar reason, the $K^+K^-\pi^0$ channel
may also be difficult to detect due to presence of $\pi^0$.
In contrast, the decay mode $K_S K^\pm \pi^\mp$  is much more
advantageous since $K_S$ can be reconstructed cleanly via
its decay to $\pi^+\pi^-$. In any event,
all these decay channels have promising observation potential
at future Super $B$ factory.

\subsection{$\eta_b$ decays to two charmed mesons}

Heavy $\eta_b$ mass opens the gate for it to
decay into charmed particles as well as light ones.
In fact, Maltoni and Polosa have suggested that $\eta_b$
decays to two charmed mesons could be the most
promising channels for observing
$\eta_b$ in Tevatron Run 2~\cite{Maltoni:2004hv}.
They first perform a perturbative calculation for $\eta_b$
decay to the inclusive $c\overline{c}g$ state,
\bqa
{\rm Br} [\eta_b\to c\overline{c} g]&=& 1.5^{+0.8}_{-0.4}
\,\%\,,
\label{MP:ccbarg}
\eqa
then interpret this value as an upper limit for the ratios of the
exclusive decays to double charmed mesons.  MP continue to
argue that, the exclusive decays to $D^* \overline{D}^{(*)}$ (the
charge conjugate is implicitly implied) may saturate the inclusive
charmful decay ratio, and subsequently estimate $10^{-3}<{\rm Br}
[\eta_b\to D^*\overline{D}^{(*)}]<10^{-2}$.

In our opinion, this saturation assumption seems to be physically
unwarranted,  consequently MP's predictions for the ratios
may well be an overestimate.  Firstly, there seems no reason to expect
that the parton process $\eta_b \to c\overline{c}g$  will
be dominated by the two-body exclusive decays,
since the $g$ jet may readily hadronize in an
independent direction,  which will result in a multi-body
decay configurations.
This can be exemplified by the fact we just mentioned,
that $\eta_c$ decays to 3 pseudo-scalars have larger
branching ratios than any 2-body decay channels.
One may notice at the lowest order in $\alpha_s$,
$\eta_b$ decay to double charm mesons is also depicted by
Fig.~\ref{QCD:diagram}, with one $c\overline{c}$  pair
replaced by a $q\overline{q}$ pair.
The gluon which is on shell in the inclusive
 $\eta_b\to c\overline{c}g$ process,  now has to
convert to a light quark pair with large invariant mass,
so is highly virtual,
the corresponding ratios of
double charm mesons thus must be at least suppressed
by one factor of $\alpha_s$ and one factor of $1/m_b^2$
relative to (\ref{MP:ccbarg}).  Taking into account the
nonperturbative binding probability,
which is much less than $1$, will further suppress
the exclusive 2-body decay
rates.  Moreover, there seems also no strong reason
to believe that the binding of $c$ with $\bar q$
will necessarily be saturated by
the ground state charm meson only.

Despite the lack of an explicit pQCD calculation, one may
still proceed with some physical consideration.
The branching ratios of $\eta_b$ decays to two charm mesons
can depend on three dimensional parameters: $m_b$, $m_c$ and
$\Lambda_{\rm QCD}$.  Since the decay $\eta_b\to D_\parallel^*\overline{D}$
conserves the helicity, we thus expect the corresponding branching
fraction scales as $1/m_b^4$.  For each pair of $c$ and $\bar{q}$
to form a $D^{(*)}$ meson, there is a factor proportional to
$\Lambda_{\rm QCD}/m_c$ which characterizes the binding
probability~\cite{Braaten:2001bf}.  This is the only place where
$\Lambda_{\rm QCD}$ dependence can enter.
Therefore, we expect the following asymptotic behavior:
\bqa
{\rm Br} [\eta_b\to D^* \overline{D}]
&\sim & \alpha_s^2 \,
\left(\lQ^2\,m_c^2\over m_b^4\right)\sim 10^{-5}\,,
\label{etab:D:Dstar}
\eqa
where $m_c^2$ is inserted to make the ratio dimensionless. Notice
this value is reasonably compatible with the inclusive upper bound,
(\ref{MP:ccbarg})~\footnote{However, as will be reported in
Ref.~\cite{Charng:Jia}, the decay rate of this process vanishes if
$D$ and $D^*$ are exactly degenerate.  Therefore, the non-vanishing
decay amplitude must be induced by heavy-quark-spin-symmetry
violating effects, $\propto \Lambda_{\rm QCD}/m_c$. This indicates
the estimate (\ref{etab:D:Dstar}) may well be subject to further
suppression.}.

We next turn to the other charmful decay channel,
$\eta_b\to D_\perp^*\overline{D}_\perp$ ($\eta_b\to D\overline{D}$ is
forbidden by $P$ invariance), which violates the helicity
selection rule maximally.  Assuming again the
helicity violation is caused by the transverse momentum
of light quark in $D$ meson, the same as in
$\eta_b \to J/\psi J/\psi$, we then obtain
\bqa
{\rm Br} [\eta_b\to D^* \overline{D}^*]
&\sim & {\rm Br} [\eta_b\to D^*\overline{D}]
\,\left(\Lambda_{\rm QCD} \over m_b\right)^4
\sim \alpha_s^2 \,
\left(\lQ^6\,m_c^2\over m_b^8\right)\sim 10^{-8}\,,
\label{etab:Dstar:Dstar}
\eqa
where  $q_\perp\sim \lQ= 500$ MeV is used.
Rather small branching ratio renders this decay mode
not so useful for detecting $\eta_b$.
Nevertheless, it might be possible
that a large prefactor may arise in the actual calculation, like
what occurs in $\eta_b \to J/\psi J/\psi$
(see footnote~\ref{prefactor}), so
that the actual ratio may be somewhat larger than this naive
estimate. In any event, a reliable pQCD calculation of the decay
rates for $\eta_b\to D^* \overline{D}^{(*)}$ will be
helpful.

Based on the previous discussions about 3-body decay of $\eta_b$,
one may wish that the branching ratio for $\eta_b\to D
\overline{D}\pi$ might be as large as $10^{-4}$,
similar to that for $\eta_b\to K \overline{K}\pi$.
However, this decay mode may not be as competitive as
$K_SK^\pm\pi^\mp$ mode, since $D$ meson does not possess
a clean signature comparable with $K_S$.

To summarize, the $\eta_b\to D^* \overline{D}$ channel,
with an expected branching ratio $\sim 10^{-5}$,
may be regarded as a valuable searching mode.
The $D^{*0}\overline{D}^0$ channel (charge conjugate state
implicitly included)
may not be so useful, since $D^{*0}$
predominantly decays to $D^0\pi^0$ and $D^0\gamma$,
where neither $\pi^0$ nor $\gamma$ can be cleanly tagged
in hadron collision environment.
In contrast, it is more advantageous to use
$D^{*+} D^-$ mode as a trigger.
$D^{*+}$ can be tagged through its decay to
$D^0\pi^+$, subsequently $D^0$ may be reconstructed from
$K^-\pi^+$, while  $D^-$ can be tagged through its decay
to $K^+\pi^-\pi^-$. It is worth emphasizing that due to the proximity
of $D^{*+}$ mass to the sum of masses of $D^0$ and $\pi^+$,
$D^{*+}$ can be cleanly identified with a rather narrow peak
in $D\pi$ invariant mass spectrum.
Using the measured branching ratios
${\rm Br}[D^{*+}\to D^0\pi^+]\approx 70\%
$, ${\rm Br}[D^0\to K^-\pi^+]\approx 4\%
$ and ${\rm Br}[D^-\to K^+\pi^-\pi^-]\approx 10\%
$~\cite{Yao:2006px}, we estimate ${\rm Br}[\eta_b \to D^{*+}D^-\to
K^+ K^- \pi^+ \pi^-\pi^+\pi^-]
\sim 10^{-8}$. There are roughly $O(1)$ produced events
in Tevatron Run 1, about $O(10^2)$ produced events in Run 2,
and about $O(10^4)$ produced events in one year run at LHC.
The statistics seems to be enough in the forthcoming hadron
collider program for observing $\eta_b$ through this decay mode,
provided that the signal events are not swallowed by
the possibly large combinatorial backgrounds.

\section{Summary and Outlook}
\label{summary}

The major motif of this work is to clarify a controversy about
whether double $J/\psi$ can be a useful decay mode to detect
$\eta_b$ in Tevatron Run 2. We have shown this process is subject to
large kinematical suppression due to the maximal violation of the
helicity selection rule. By an explicit pQCD calculation based on
NRQCD, we predict the corresponding branching ratio to be only of
order $10^{-8}$, thus making the search for $\eta_b$ through this
mode rather unrealistic in Tevatron Run 2. Nevertheless, we
anticipate that at LHC, the observation potential of this decay mode
may not be so pessimistic,  if experimentalists can find a good way
to suppress the rather copious QCD background events.

The large mass of $\eta_b$ renders any of its exclusive decay
channels in general very small.  To provide some useful guidance for
experimental search for this elusive particle, it is valuable to
identify those decay modes with relatively large branching fractions
and also with clean signature. To make progress along this
direction, in the following we outline some issues which we think
may deserve further studies:

\begin{enumerate}

\item{
Stimulated by rather large QCD radiative correction to exclusive
double charmonium production at $B$
factory~\cite{Zhang:2005cha,Gong:2007db}, one may inquire how large
the effect of the next-to-leading order QCD correction to $\eta_b\to
J/\psi J/\psi$ is. It turns out that, at NLO in $\alpha_s$ for this
process, the helicity selection rule can be violated by finite charm
mass, consequently one obtains the non-vanishing result even at the
the leading order in velocity expansion~\cite{Gong:Jia:Wang}.  At
the amplitude level, the ratio of the radiative correction piece to
the relativistic correction piece as considered in this work,
(\ref{amplitude:strong}),  is about ${\alpha_s\over \pi}:\langle
v^2\rangle_{J/\psi}\sim {\cal O}(1)$, which implies both effects are
equally important,  and the more accurate prediction will crucially
depend on their relative phase.   To this end,  a precise
determination of the quantity $\langle v^2\rangle_{J/\psi}$, in
particular its sign, would be helpful}.

\item{The process $\eta_b\to K^* \overline{K}+{\rm c.c.}$,
favored by the helicity selection rule but at the same time
violating U-spin symmetry,  is estimated to have a branching
fraction $\sim 10^{-6}$-$10^{-7}$.  It might be worthwhile if an
actual pQCD calculation which implements $m_s-m_d$ difference,  can
be carried out in the light cone expansion scheme,  to compare with
this rough estimate.  Moreover, the eventual experimental sighting
of $\eta_c\to K^* \overline{K}$ in charmonium factory like BES III
will definitely enrich our understanding toward this class of
helicity-conserving yet flavor SU(3) violating quarkonium decay
processes. }

\item{
The individual decay modes of $\eta_c$ with largest branching
ratios are $\eta_c\to K \overline{K}\pi$, $\eta\pi\pi$ and
$\eta^\prime \pi\pi$.  Stimulated by this experimental fact, one
may hope that the 3-body decay channels of $\eta_b$,
such as $K_SK^\pm\pi^\mp$,
with an expected branching ratio of order $10^{-4}$, might be a
potentially useful searching mode for $\eta_b$ in current and
forthcoming hadron collider programs, if it can survive from
the copious combinatorial background events.
This mode will definitely have
promising potential to be observed
in the prospective Super $B$ factory.}

\item{
Another helicity-conserving decay process, $\eta_b\to D^*
\overline{D}+{\rm c.c.}$, with an expected branching ratio of order
$10^{-5}$, may also be well worth searching for experimentally. It
is also profitable to carry out a concrete calculation of this
exclusive double charm decay process from pQCD scheme,  but one may
be obliged to incorporate the heavy-quark-spin-symmetry violating
effect~\cite{Charng:Jia}.}

\item{
In this work we haven't discussed the possibility of observing
$\eta_b$ through its decay to baryon pair, such as $\eta_b \to
p\overline{p}$. If the corresponding ratio is not too small, this is
potentially a good searching mode thanks to relatively fewer
baryonic background events in hadron collision
experiments~\footnote{I thank C.~Z.~Yuan for suggesting this
possibility.}. Experimentally $\eta_c \to p\bar p$ is observed to
have a branching ratio of order $10^{-3}$~\cite{Yao:2006px}.
However, one should be aware that this process also violates the
helicity selection rule, and the available pQCD prediction, when
taking into account the nonzero light quark mass but still in the
collinear quark configuration, is still far smaller than the
measured value~\cite{Anselmino:1992jd}, therefore some
nonperturbative mechanism needs to be called for to explain this
discrepancy~\cite{Feldmann:2000hs}.  Because of rather heavy mass of
$\eta_b$, one may hope pQCD framework should provide a reliable
prediction for $\eta_b \to p\overline{p}$. It will be valuable if a
more systematic calculation can be carried out. }

\end{enumerate}


{\noindent \large \it \underline{Note added.}} $\;\;$  While the
revised version of the manuscript is to be submitted,  there appears
an eprint in arXiv by \textsc{Babar}
collaboration~\cite{Aubert:2008vj}, which claims the first
unambiguous discovery of $\eta_b$ through the hindered $M1$
transition process $\Upsilon(3S)\to \eta_b\gamma$. It is interesting
to note the rather precisely measured $\eta_b$ mass,
$9388.9^{+3.1}_{-2.3}(stat)\pm2.7(syst)$ MeV,  seems not compatible
with the predictions from most potential models as well as the
weakly-coupled potential NRQCD, instead consistent with the lattice
QCD prediction within error~\cite{Gray:2005ur}.
\\

{\noindent \large \it \underline{Acknowledgment.}} $\;\;$ I wish to
thank Augustine Chen,  Tao Huang, Chang-Zheng Yuan, Qiang Zhao and
Bing-Song Zou for helpful discussions, Yeo-Yie Charng, Bin Gong and
Jian-Xiong Wang for enjoyable collaborations on topics closely
related to this work. This research was partially supported by
National Natural Science Foundation of China under Grant
No.~10605031.



\begin{thebibliography}{99}



\bibitem{Brambilla:2004wf}
For a comprehensive review and more references,
see N.~Brambilla {\it et al},
CERN-2005-005 (CERN, Geneva, 2005)
[arXiv:hep-ph/0412158].


\bibitem{Eichten:1994gt}
  E.~J.~Eichten and C.~Quigg,
  Phys.\ Rev.\ D {\bf 49}, 5845 (1994)
  [arXiv:hep-ph/9402210].


\bibitem{Narison:1995tw}
  S.~Narison,
  Phys.\ Lett.\ B {\bf 387}, 162 (1996)
  [arXiv:hep-ph/9512348].



\bibitem{Lengyel:2000dk}
  V.~Lengyel, Yu.~Fekete, I.~Haysak and A.~Shpenik,
  Eur.\ Phys.\ J.\ C {\bf 21}, 355 (2001)
  [arXiv:hep-ph/0007084].


\bibitem{Brambilla:2001fw}
  N.~Brambilla, Y.~Sumino and A.~Vairo,
  Phys.\ Lett.\ B {\bf 513}, 381 (2001)
  [arXiv:hep-ph/0101305].



\bibitem{Barnes:2001rt}
  T.~Barnes,
  arXiv:hep-ph/0103142.



\bibitem{Liao:2001yh}
  X.~Liao and T.~Manke,
  Phys.\ Rev.\ D {\bf 65}, 074508 (2002)
  [arXiv:hep-lat/0111049].


\bibitem{Ebert:2002pp}
  D.~Ebert, R.~N.~Faustov and V.~O.~Galkin,
  Phys.\ Rev.\ D {\bf 67}, 014027 (2003)
  [arXiv:hep-ph/0210381].



\bibitem{Recksiegel:2003fm}
  S.~Recksiegel and Y.~Sumino,
  Phys.\ Lett.\ B {\bf 578}, 369 (2004)
  [arXiv:hep-ph/0305178].

\bibitem{Kniehl:2003ap}
  B.~A.~Kniehl, A.~A.~Penin, A.~Pineda, V.~A.~Smirnov and M.~Steinhauser,
  Phys.\ Rev.\ Lett.\  {\bf 92}, 242001 (2004)
  [arXiv:hep-ph/0312086].



\bibitem{Gray:2005ur}
  A.~Gray, I.~Allison, C.~T.~H.~Davies, E.~Gulez, G.~P.~Lepage, J.~Shigemitsu and M.~Wingate,
  Phys.\ Rev.\ D {\bf 72}, 094507 (2005)
  [arXiv:hep-lat/0507013].



\bibitem{Heister:2002if}
  A.~Heister {\it et al.}  [ALEPH Collaboration],
  Phys.\ Lett.\ B {\bf 530}, 56 (2002)
  [arXiv:hep-ex/0202011].


\bibitem{Levtchenko:2004ku}
  M.~Levtchenko  [L3 Collaboration],
  Nucl.\ Phys.\ Proc.\ Suppl.\  {\bf 126}, 260 (2004).



\bibitem{Abdallah:2006yg}
  J.~Abdallah  [DELPHI Collaboration],
  Phys.\ Lett.\ B {\bf 634}, 340 (2006)
  [arXiv:hep-ex/0601042].


\bibitem{Artuso:2004fp}
  M.~Artuso {\it et al.}  [CLEO Collaboration],
  Phys.\ Rev.\ Lett.\  {\bf 94}, 032001 (2005)
  [arXiv:hep-ex/0411068].


\bibitem{Braaten:2000cm}
  E.~Braaten, S.~Fleming and A.~K.~Leibovich,
  Phys.\ Rev.\ D {\bf 63}, 094006 (2001)
  [arXiv:hep-ph/0008091].

\bibitem{Groom:2000in}
  D.~E.~Groom {\it et al.}  [Particle Data Group],
  Eur.\ Phys.\ J.\ C {\bf 15}, 1 (2000).



\bibitem{Yao:2006px}
  W.~M.~Yao {\it et al.}  [Particle Data Group],
  J.\ Phys.\ G {\bf 33}, 1 (2006).



\bibitem{Tseng:2003md}
  J.~Tseng  [CDF collaboration],
FERMILAB-CONF-02-348-E,
{\it Presented at 5th International Conference on
Quark Confinement and the Hadron Spectrum, Gargnano,
Brescia, Italy, 10-14 Sep 2002}.


\bibitem{Maltoni:2004hv}
  F.~Maltoni and A.~D.~Polosa,
  Phys.\ Rev.\ D {\bf 70}, 054014 (2004)
  [arXiv:hep-ph/0405082].



\bibitem{Anselmino:1990vs}
  M.~Anselmino, F.~Murgia and F.~Caruso,
  Phys.\ Rev.\ D {\bf 42}, 3218 (1990).


\bibitem{jia:1998}
  Y.~Jia and G.~D.~Zhao,
  High\ Energy\  Phys.\  Nucl.\ Phys. {\bf 23}, 765
  (1999) (in Chinese);
 \\
  Y.~Jia, {\it Some exclusive strong decay processes of heavy quarkonia},
  MS thesis, (Peking U., 1998).


\bibitem{Feldmann:2000hs}
  T.~Feldmann and P.~Kroll,
  Phys.\ Rev.\ D {\bf 62}, 074006 (2000)
  [arXiv:hep-ph/0003096].



\bibitem{Zhou:2005fc}
  H.~Q.~Zhou, R.~G.~Ping and B.~S.~Zou,
  Phys.\ Rev.\ D {\bf 71}, 114002 (2005).



\bibitem{Braaten:2002fi}
  E.~Braaten and J.~Lee,
  Phys.\ Rev.\ D {\bf 67}, 054007 (2003)
  [Erratum-ibid.\ D {\bf 72}, 099901 (2005)]
  [arXiv:hep-ph/0211085].


\bibitem{Bodwin:2002fk}
  G.~T.~Bodwin, J.~Lee and E.~Braaten,
  Phys.\ Rev.\ Lett.\  {\bf 90}, 162001 (2003)
  [arXiv:hep-ph/0212181].


\bibitem{Bodwin:2002kk}
  G.~T.~Bodwin, J.~Lee and E.~Braaten,
  Phys.\ Rev.\ D {\bf 67}, 054023 (2003)
  [Erratum-ibid.\ D {\bf 72}, 099904 (2005)]
  [arXiv:hep-ph/0212352].



\bibitem{Liu:2002wq}
  K.~Y.~Liu, Z.~G.~He and K.~T.~Chao,
  Phys.\ Lett.\ B {\bf 557}, 45 (2003)
  [arXiv:hep-ph/0211181].


\bibitem{Hagiwara:2003cw}
  K.~Hagiwara, E.~Kou and C.~F.~Qiao,
  Phys.\ Lett.\ B {\bf 570}, 39 (2003)
  [arXiv:hep-ph/0305102].

\bibitem{Zhang:2005cha}
  Y.~J.~Zhang, Y.~j.~Gao and K.~T.~Chao,
  Phys.\ Rev.\ Lett.\  {\bf 96}, 092001 (2006)
  [arXiv:hep-ph/0506076].

\bibitem{Gong:2007db}
  B.~Gong and J.~X.~Wang,
  Phys.\ Rev.\  D {\bf 77}, 054028 (2008)
  [arXiv:0712.4220 [hep-ph]].


\bibitem{Brodsky:1981kj}
  S.~J.~Brodsky and G.~P.~Lepage,
  Phys.\ Rev.\ D {\bf 24}, 2848 (1981).


\bibitem{Chernyak:1983ej}
  V.~L.~Chernyak and A.~R.~Zhitnitsky,
  Phys.\ Rept.\  {\bf 112}, 173 (1984);

  V.~L.~Chernyak and A.~R.~Zhitnitsky,
  Nucl.\ Phys.\ B {\bf 201}, 492 (1982)
  [Erratum-ibid.\ B {\bf 214}, 547 (1983)].


\bibitem{Braguta:2005gw}
  V.~V.~Braguta, A.~K.~Likhoded and A.~V.~Luchinsky,
  Phys.\ Rev.\ D {\bf 72}, 094018 (2005)
  [arXiv:hep-ph/0506009].



\bibitem{Zhao:2006cx}
  Q.~Zhao,
  Phys.\ Lett.\ B {\bf 636}, 197 (2006)
  [arXiv:hep-ph/0602216].


\bibitem{Peskin:1979va}
  M.~E.~Peskin,
  Nucl.\ Phys.\ B {\bf 156}, 365 (1979);

G.~Bhanot and M.~E.~Peskin,
  Nucl.\ Phys.\ B {\bf 156}, 391 (1979).


\bibitem{Bodwin:1994jh}
  G.~T.~Bodwin, E.~Braaten and G.~P.~Lepage,
  Phys.\ Rev.\  D {\bf 51}, 1125 (1995)
  [Erratum-ibid.\  D {\bf 55}, 5853 (1997)]
  [arXiv:hep-ph/9407339].

\bibitem{Bodwin:2002hg}
  G.~T.~Bodwin and A.~Petrelli,
  Phys.\ Rev.\ D {\bf 66}, 094011 (2002)
  [arXiv:hep-ph/0205210].


\bibitem{Bodwin:1996tg}
G.~T.~Bodwin, D.~K.~Sinclair and S.~Kim,
Phys.\ Rev.\ Lett.\  {\bf 77}, 2376 (1996)
[arXiv:hep-lat/9605023].


\bibitem{Gremm:1997dq}
  M.~Gremm and A.~Kapustin,
  Phys.\ Lett.\ B {\bf 407}, 323 (1997)
  [arXiv:hep-ph/9701353].

\bibitem{Brambilla:2002nu}
  N.~Brambilla, D.~Eiras, A.~Pineda, J.~Soto and A.~Vairo,
  Phys.\ Rev.\  D {\bf 67}, 034018 (2003)
  [arXiv:hep-ph/0208019].

\bibitem{Eidemuller:2002wk}
  M.~Eidemuller,
  Phys.\ Rev.\  D {\bf 67}, 113002 (2003)
  [arXiv:hep-ph/0207237].


\bibitem{Bodwin:2006dn}
  G.~T.~Bodwin, D.~Kang and J.~Lee,
  Phys.\ Rev.\ D {\bf 74}, 014014 (2006)
  [arXiv:hep-ph/0603186].

\bibitem{Braguta:2006wr}
  V.~V.~Braguta, A.~K.~Likhoded and A.~V.~Luchinsky,
  Phys.\ Lett.\  B {\bf 646}, 80 (2007)
  [arXiv:hep-ph/0611021].


\bibitem{Braaten:2001bf}
  E.~Braaten, Y.~Jia and T.~Mehen,
  Phys.\ Rev.\ D {\bf 66}, 034003 (2002)
  [arXiv:hep-ph/0108201].


\bibitem{Charng:Jia}
Y.-Y.~Charng and Y.~Jia, work in progress.


\bibitem{Barger:1995vx}
  V.~D.~Barger, S.~Fleming and R.~J.~N.~Phillips,
  Phys.\ Lett.\ B {\bf 371}, 111 (1996)
  [arXiv:hep-ph/9510457].


\bibitem{Qiao:2002rh}
  C.~F.~Qiao,
  Phys.\ Rev.\ D {\bf 66}, 057504 (2002)
  [arXiv:hep-ph/0206093].



\bibitem{Ablikim:2005yi}
  M.~Ablikim {\it et al.}  [BES Collaboration],
  Phys.\ Rev.\ D {\bf 72}, 072005 (2005)
  [arXiv:hep-ex/0507100].

\bibitem{Feldmann:1998vh}
  T.~Feldmann, P.~Kroll and B.~Stech,
  Phys.\ Rev.\  D {\bf 58}, 114006 (1998)
  [arXiv:hep-ph/9802409].

\bibitem{Hewett:2004tv}
  J.~Hewett {\it et al.},
  arXiv:hep-ph/0503261.


\bibitem{Gong:Jia:Wang}
B.~Gong, Y.~Jia and J.-X.~Wang, in preparation.

\bibitem{Anselmino:1992jd}
  M.~Anselmino, R.~Cancelliere and F.~Murgia,
  Phys.\ Rev.\ D {\bf 46}, 5049 (1992).

\bibitem{Aubert:2008vj}
B.~Aubert  [The BABAR Collaboration],
arXiv:0807.1086 [hep-ex].


\end{thebibliography}
\end{document}